\documentclass[fleqn,usenatbib]{mnras}

\usepackage[T1]{fontenc}

\DeclareRobustCommand{\VAN}[3]{#2}
\let\VANthebibliography\thebibliography
\def\thebibliography{\DeclareRobustCommand{\VAN}[3]{##3}\VANthebibliography}

\usepackage{float}
\usepackage{graphicx}	
\usepackage{amsmath}	
\usepackage{amssymb}	

\newcommand{\email}[1]{\href{mailto:#1}{\tt #1}}
\newcommand{\B}[1]{\beta_\text{#1}}

\title[Optical darkness in short GRBs]{Optical darkness in short-duration $\gamma$-ray bursts}

\author[Gobat, van der Horst, \& Fitzpatrick]{
Caden Gobat$^{1,2,}$\thanks{E-mail: \email{cgobat@gwu.edu}},
Alexander J. van der Horst$^{1}$, David Fitzpatrick$^{3,4}$
\\
$^1$Department of Physics, George Washington University, 725 21st St NW, Washington, DC 20052, U.S.A.\\ 
$^2$Department of Space Operations, Southwest Research Institute, 1050 Walnut Street, Suite 300, Boulder, CO 80302, U.S.A.\\
$^3$Department of Aerospace Engineering Sciences, University of Colorado Boulder, 3775 Discovery Dr, Boulder, CO 80303, U.S.A.\\
$^4$Department of Physics, Georgetown University, 37th \& O St NW, Washington, DC 20007, U.S.A.
}

\date{Accepted 2023 April 13. Received 2023 April 10; in original form 2023 January 18} 

\pubyear{2023}

\usepackage{newtxtext,newtxmath}

\begin{document}
\label{firstpage}
\pagerange{\pageref{firstpage}--\pageref{lastpage}}
\maketitle

\begin{abstract}
Gamma-ray bursts categorically produce broadband afterglow emission, but in some cases, emission in the optical band is dimmer than expected based on the contemporaneously observed X-ray flux. This phenomenon, aptly dubbed ``optical darkness'', has been studied extensively in long GRBs (associated with the explosive deaths of massive stars), with possible explanations ranging from host environment extinction to high redshift to possibly unique emission mechanisms. However, investigations into optical darkness in short GRBs (associated with the mergers of compact object binaries) have thus far been limited. This work implements a procedure for determining the darkness of GRBs based on spectral indices calculated using temporally-matched \textit{Swift}-XRT data and optical follow-up observations; presents a complete and up-to-date catalog of known short GRBs that exhibit optical darkness; and outlines some of the possible explanations for optically dark short GRBs. In the process of this analysis, we developed versatile and scalable data processing code that facilitates reproducibility and reuse of our pipeline. These analysis tools and resulting complete sample of dark short GRBs enable a systematic statistical study of the phenomenon and its origins, and reveal that optical darkness is indeed quite rare in short GRBs, and highly dependent on observing response time and observational effects.
\end{abstract}

\begin{keywords}
gamma-ray bursts -- neutron star mergers -- dust, extinction -- methods: observational
\end{keywords}



\section{Introduction}

Gamma-ray bursts (GRBs) are some of the brightest transient astrophysical phenomena observed in the Universe. The prevailing hypothesis is that GRBs are the products of two main classes of progenitors: collapsar events, resulting from end-of-life core collapse in supermassive stars \citep{1993ApJ...405..273W}, or compact object (neutron star-neutron star or possibly neutron star-black hole) binary mergers \citep{1989Natur.340..126E}. Observationally, GRBs are generally split into sub-populations corresponding to these two progenitor types on the basis of the burst's $T_{90}$ duration \citep[see][]{1993ApJ...413L.101K}, the time it takes for it to emit 90\% of its gamma-ray radiation.
Short GRBs, typically associated with binary neutron star (BNS) mergers,\footnote{Observational confirmation that BNS mergers are linked to short GRBs came with the simultaneous observation of kilonova AT2017gfo, GRB\,170817A, and gravitational wave event GW170817 \citep{2017ApJ...848L..12A}.}
are usually taken to be those with $T_{90}\text{ \footnotesize$\lesssim$ }2$ seconds, and long GRBs, resulting from core collapse events, are those with $T_{90}\text{ \footnotesize$\gtrsim$ }2$ seconds. When available, other metrics can be used to classify GRBs as well, such as spectral hardness and luminosity (short GRBs are typically spectrally harder but less luminous than long ones). For an in-depth review of gamma-ray bursts, see \citet{2009ARA&A..47..567G}.

Categorically, after the initial burst, called prompt emission, GRBs produce an afterglow, which refers to a period of fading multi-wavelength emission that lasts from hours to days and sometimes years after the initial event. In 1997, \citeauthor{1997Natur.386..686V} announced the discovery of the first optical counterpart to a GRB, and transient optical afterglows have been identified from many GRBs since. Shortly after the initial discovery, however, \citet{1998ApJ...493L..27G} reported on the discovery of a GRB with no detectable optical afterglow. Since then, observers have found that a fraction of all GRB afterglows exhibit a phenomenon known as optical darkness, where the afterglow as a whole is clearly detected (typically in the X-rays) and yet much dimmer than expected or not present at all in the optical band. This phenomenon has been observed in both long and short GRBs \citep{2011A&A...526A..30G}, with a number of proposed explanations \citep{2001A&A...369..373F}. However, the implications of optical darkness differ somewhat for the two different classes of progenitor. 

The massive stars associated with long GRBs live fast and die young, while the compact object binaries that give rise to short GRBs must be old enough for both members to have gone supernova and turned into neutron stars, and then orbit each other for long enough to spiral inwards and collide. One possible cause of optical darkness in some GRBs is that emission has been redshifted towards the infrared due to cosmological distance and the expansion of the Universe, but this would imply that the event occurred long ago, when the Universe was relatively young. This explanation makes sense for long GRBs, which can reasonably be expected to be possible within $\sim1$ Gyr after the Big Bang. However, compact binary merger events should not be expected to occur often at this early stage of the Universe's evolution, as it is unlikely that they would have had time to form \citep{2007ApJ...665.1220Z, 2015MNRAS.448.3026W, 2019MNRAS.487.4847B, 2020A&A...634L...2S}. Another proposed explanation is host galaxy extinction \citep{2002MNRAS.330..583L, 2013MNRAS.432.1231C, 2015MNRAS.449.2919L}, which refers to the possibility that gas and dust within the galaxy where the GRB occurred blocks light in the optical band. Again, this is rational for long GRBs, which occur in regions of star-forming activity amidst environments of dense gas and dust. However, this is not so universally applicable for short GRBs, whose progenitors often travel far away from where the stars formed \citep[e.g.,][]{2010ApJ...722.1946B,2022MNRAS.515.4890O}, and are not as predictably found in these kinds of regions. A final explanation is that optical darkness is an intrinsic property of certain GRBs: some simply might not emit as much optical light as others. However, this unique physics explanation is disfavored, at least for long GRBs \citep{2005ApJ...624..868R}. Partly as a result of these contrasts, and partly due to the overall difference in afterglow brightness of long and short GRBs, optical darkness in long GRBs has been studied much more extensively than in short ones. In this work we focus on short GRBs in order to better understand the environments in which they form and the histories of the systems that produce them.

There are two main criteria in the literature for determining if a burst is optically dark, and both depend on the optical-to-X-ray spectral power-law index, $\B{ox}$, which provides a metric for the relative flux intensity at X-ray versus optical frequencies. In general, a low value of $\B{ox}$ means that the power-law slope between the optical and X-rays is shallow, or perhaps even positive (we utilize the sign convention of $F_\nu\propto\nu^{-\beta}$, meaning $\beta<0$ implies a positive slope and consequently higher emission in the X-rays than in the optical). The first method \citep{2004ApJ...617L..21J} defines optical darkness with a cutoff of $\B{ox}<0.5$, which is derived from the assumption that the number distribution of electron Lorentz factors in the burst outflow (described by $n_e(\gamma) \propto \gamma^{-p}$) is limited by $p>2$. Integrating the synchrotron emission produced by such an electron distribution results in a lower limit of $\beta>0.5$ on the broadband spectrum, meaning violations of this should be considered abnormal (i.e., ``dark''). However, cases of $p<2$ have been found \citep[e.g.,][]{2001A&A...374..382M, 2001ApJ...563..592S}, so this assumption is not universally valid.

The second method \citep{2009ApJ...699.1087V} incorporates a burst's X-ray spectral information, which is routinely available thanks to the rapid follow-up capabilities of the Neil Gehrels \textit{Swift} Observatory (hereafter \textit{Swift}) and its X-ray Telescope \citep[\textit{Swift}--XRT,][]{2005SSRv..120..165B}. If the spectral index in the X-ray regime ($\B{x}$) is known, it can be compared to the optical-to-X-ray spectral index. \citeauthor{2009ApJ...699.1087V} define a burst as optically dark if $\B{ox}<\B{x}-0.5$, allowing for the possibility that $p<2$. Given current physical models for GRB afterglow emission mechanisms, and assuming that the optical and X-ray emission are part of the same broadband spectrum, all bursts should in theory lie in the region $\B{x}-0.5<\B{ox}<\B{x}$, meaning that if $\B{ox}$ is below this range, the burst is optically dark.

This work implements a scalable pipeline for determining darkness using both of these methods, with special care given to sample completeness and systematic procedures for retrieval of X-ray spectra and light curves, as well as ultraviolet (UV), optical, and near-infrared (nIR) follow-up observations. Using this, we present a complete catalog of optically-dark short GRBs since \textit{Swift}'s launch in 2004 through the end of 2021, which offers insight into the mechanisms and possible causes of the phenomenon.

In Section~\ref{sec:methods}, we present the methodology, sample selection and pipeline we developed for this work. We show the results in  Section~\ref{sec:results}, and discuss the implications for the entire short GRB population and a few interesting cases in Section~\ref{sec:discussion}. Section~\ref{sec:conclusion} summarizes and concludes this paper.

\section{Methodology \& Pipeline}\label{sec:methods}

To determine $\B{ox}$ for a given GRB, we first require the existence of time-resolved flux data in both the optical/nIR/UV and X-ray bands for comparison. Multi-wavelength data must also be relatively coincident in time to make a valid comparison, due to the afterglow's rapid temporal decay. We also require existing X-ray spectral fits, given that the \citet{2009ApJ...699.1087V} criterion for optical darkness depends on $\B{x}$, the X-ray spectral index.

To accomplish this, we implemented a mostly automated data reduction and analysis pipeline in Python, with several distinct, consecutive sections. X-ray flux data and spectral information are retrieved automatically from online repositories as a part of this process, while optical data is compiled manually. The code searches for temporal matches between these two data sets for each burst within a user-defined tolerance, and calculates $\B{ox}$ using appropriately converted fluxes that have been corrected for Galactic extinction.

\subsection{Sample definition and data collection}\label{sec:sample}

Our sample definition starts from the master-level \href{https://swift.gsfc.nasa.gov/archive/grb_table/}{\textit{Swift} GRB table}, with updated $T_{90}$ values pulled from the \href{https://swift.gsfc.nasa.gov/results/batgrbcat/}{\textit{Swift} Burst Alert Telescope catalog pages}. From there, we select all GRBs up through the end of 2021 that meet one or more of the following criteria: \begin{itemize}
    \item $T_{90}\leq2$ seconds;
    \item present in the sample of \citet{2015ApJ...815..102F}, whose criteria were:
    \begin{itemize}
        \item occurrence between November 2004 and March 2015, and
        \item $T_{90}\lesssim2$ seconds, with exceptions made for GRBs 050724A, 090607 and 100213A (which have $T_{90}$ between 2.5 and 3 seconds) on the basis of spectral lag/hardness ratio, and
        \item follow-up observations in the X-ray, optical, near-infrared, or radio bands are available.
    \end{itemize}
    \item present in the sample of \citet{2021ApJ...916...89R}, whose criteria were:
    \begin{itemize}
        \item occurrence between 2005 and 2020, and
        \item detected by \textit{Swift}--BAT, and
        \item $T_{90}\lesssim2$ seconds, or classified as short in the \textit{Swift}--BAT catalog \citep{2016ApJ...829....7L}.
    \end{itemize}
    \item GCN Circular announcement(s)\footnote{\url{https://gcn.gsfc.nasa.gov/selected.html}} for the burst identify it as short.\footnote{The process for identifying these bursts was partially automated, and GCN Circular scraping code can be found on GitHub at \href{https://github.com/cgobat/dark-GRBs/blob/master/GCNs.ipynb}{\texttt{cgobat/dark-GRBs/GCNs.ipynb}} and \href{https://github.com/cgobat/dark-GRBs/blob/master/catalog generator.ipynb}{\texttt{cgobat/dark-GRBs/catalog generator.ipynb}}}
\end{itemize}

This results in a master table with a list of GRB identifiers as well as basic information about the bursts.

The calculation of $\B{ox}$ requires contemporaneous flux measurements in the X-ray and optical bands. The UK \textit{Swift} Science Data Center \citep[UKSSDC;][]{2007A&A...469..379E, 2009MNRAS.397.1177E} provides X-ray data taken by \textit{Swift}--XRT. To collect these data, we wrote custom retrieval scripts to query the online repository to scrape each burst's spectrum page\footnote{\url{https://www.swift.ac.uk/xrt_spectra/}} for its X-ray photon index, $\Gamma$ (from which we get the X-ray spectral index using the relationship $\Gamma=1+\B{x}$), and intrinsic column density, $N_H$. We parse the lightcurve page\footnote{\url{https://www.swift.ac.uk/xrt_curves/}} to obtain an X-ray flux time series, and the XRT Live Catalog pages\footnote{\url{https://www.swift.ac.uk/xrt_live_cat/}} for fitted temporal indices ($\alpha$) and corresponding light curve break times.

The majority of our optical, nIR, and UV data were parsed and compiled manually from GCN Circular announcements of follow-up observation results for GRBs of interest, plus data presented in peer-reviewed publications. In recent years, GRB localizations published through the GCN have enabled increasingly frequent and rapid follow-up observations by telescopes around the world. By compiling data from the GCN Circulars for each of the bursts in our sample, we compile a table of magnitudes of optical detections, upper limits, and associated errors, as well as the observation time and observing instrument/band. We supplement these data with the short GRB optical observations presented in \citet{2015ApJ...815..102F} and \citet{2021ApJ...916...89R}.

\subsection{X-ray data processing}

The UKSSDC provides \textit{Swift}-XRT light curves for GRBs in units of integrated flux across the entire XRT 0.3--10~keV band. To compare fluxes at specific wavelengths, it is necessary to convert these integrated fluxes into spectral flux densities (i.e., units of Jy or similar). Within the X-ray band, we assume the afterglow spectrum (as a function of frequency, $\nu$) to be described by a single power law with spectral index $\B{x}$. This yields the following relation:
\begin{equation}\label{eq:integral}
    F_\text{x} = \int_{0.3\text{ keV}}^{10\text{ keV}} F_E\ dE = A \int_{7.3\cdot10^{16}\text{ Hz}}^{2.4\cdot10^{18}\text{ Hz}} \nu^{-\B{x}}\ d\nu
\end{equation}
where $F_\text{x}$ is the bolometric X-ray flux value from \textit{Swift}-XRT and $A$ is a scaling coefficient that accounts for intrinsic X-ray luminosity and distance to the burst. The value of $A$ can be determined via the analytical solution to Eq. (\ref{eq:integral}), which is
\begin{equation}
    F_\text{x} = A\cdot\left.\begin{cases} \ln(\nu) & \text{if }\B{x} = 1 \\
    \frac{\nu^{1-\beta}}{1-\beta} & \text{otherwise} \end{cases}\right|_{\nu=7.3\cdot10^{16}\text{ Hz}}^{\nu=2.4\cdot10^{18}\text{ Hz}}
\end{equation}

We can thus evaluate $A\nu^{-\B{x}}$ at $\nu = 10^\frac{\log(10)+\log(0.3)}{2}\approx1.732$~keV, the logarithmic midpoint frequency of the 0.3--10~keV range, to get $F_{\nu,\text{x}}$, the spectral flux in the logarithmic middle of the X-ray band. We perform this computation for every entry in the XRT data table (on the order of several tens of data points for each GRB) to compute a usable spectral flux for each flux data point.

\subsection{Optical data processing}

Information collected in the optical regime includes observation time, magnitude or limiting magnitude, magnitude error, observation filter and its effective wavelength $\lambda_\text{eff}$, and the Galactic reddening $E_{B-V}$ in the direction of the GRB. 
For the most part, magnitudes are reported in the AB system. Where Vega magnitudes are reported, as in the case of the \textit{Swift} UV/Optical Telescope \citep[UVOT,][]{2005SSRv..120...95R}, we apply a correction factor to translate into AB magnitudes.\footnote{AB--Vega correction offsets for \textit{Swift}-UVOT filters can be found at \url{https://swift.gsfc.nasa.gov/analysis/uvot_digest/zeropts.html}}

\begin{figure}
    \centering
    \includegraphics[width=\linewidth]{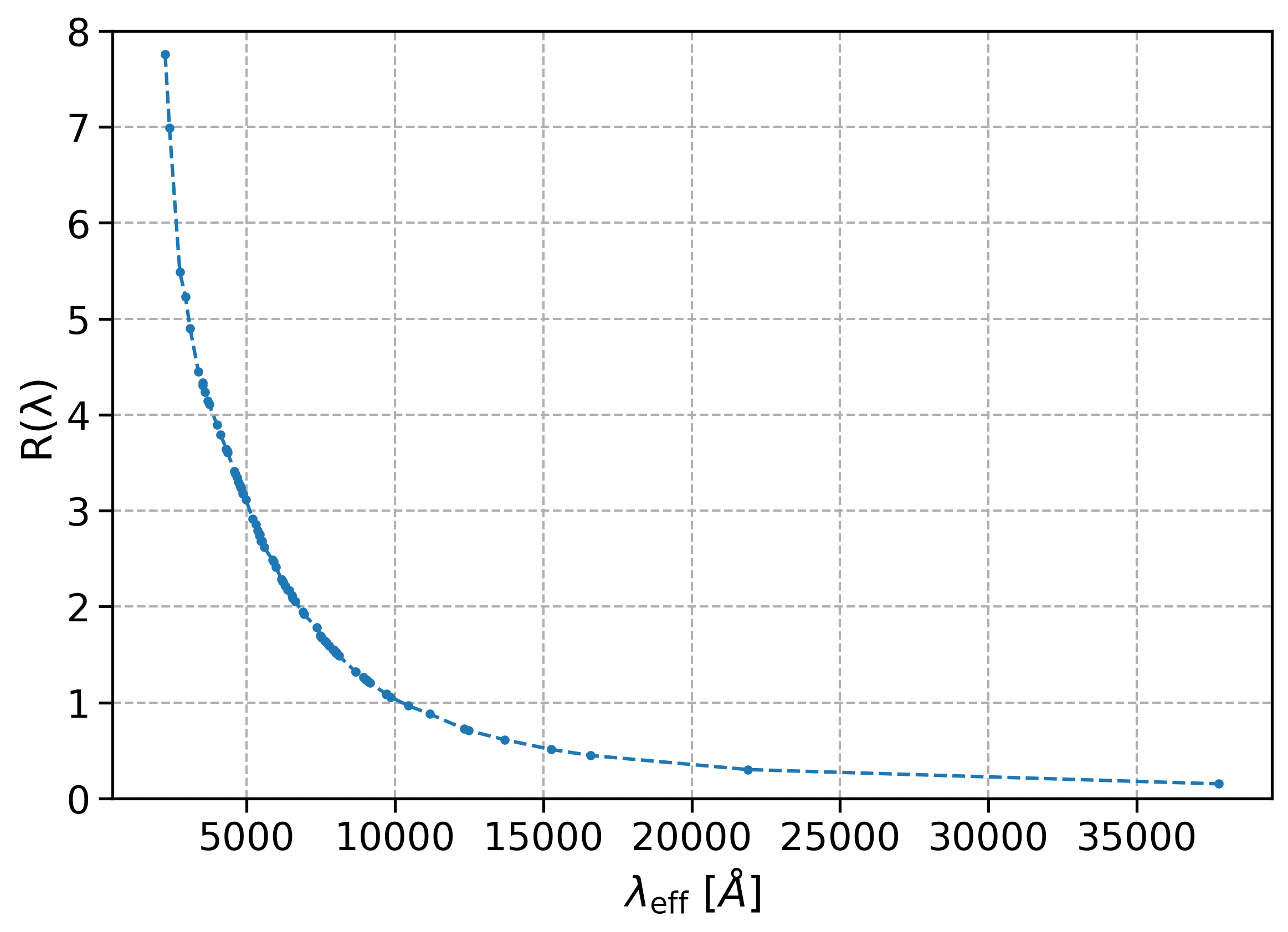}
    \caption{Relationship between an observing filter's effective wavelength and coefficient $R_\lambda$ that we use to convert between $E_{B-V}$ reddening and magnitude extinction.}
    \label{fig:Rb}
\end{figure}

The extinction at some wavelength $\lambda$ is given by $A_\lambda = R_\lambda\cdot E_{B-V}$, where $R_\lambda$ is dependent on $\lambda$ and $E_{B-V}$ is an observed property of the interstellar medium along the line-of-sight. To calculate this $R(\lambda)$ for Galactic extinction, we establish an interpolatory function using empirical extinction data provided in \citet{2011ApJ...737..103S}'s Table 6. The relationship is well-defined, as shown in Figure \ref{fig:Rb}. Knowing this and the reddening value for each GRB, we calculate the Galactic extinction and adjust our magnitude values for it. With all of this information, we then calculate a spectral flux or upper limit for the observation according to \begin{equation}\label{eq:optflux}
    F_{\nu,\text{o}} = 3631\cdot10^{-\frac{m_\lambda-A_\lambda}{2.5}}\text{ Jy}
\end{equation}

\subsection{Temporal matching \& calculation of $\B{ox}$}

With unit-compatible flux values in the X-ray and optical bands, the next step is to match these data points in time. We define the fractional temporal separation between optical observation time ($t_\text{o}$) and X-ray observation time ($t_\text{x}$) as \begin{equation}\label{eq:dt}
    dt_\% = \frac{|t_\text{o}-t_\text{x}|}{t_\text{x}}
\end{equation}
and compute this for every combination of optical and X-ray data points within each GRB, accepting any pair of data points for which $dt_\% \leq 0.2$ as a candidate match.

For each temporal match of X-ray and optical data, we calculate $\B{ox}$, the power-law slope between the two points:
\begin{equation}\label{eq:B_ox}
    \B{ox} = -\frac{\log(F_{\nu,\text{x}}/F_{\nu,\text{o}})}{\log(\nu_\text{x}/\nu_\text{o})}
\end{equation}

Because this calculation assumes exactly contemporaneous flux observations in the two bands, we must also account for some additional error due to time-dependent afterglow decay. If the afterglow is fading rapidly, even a small difference in time could correspond to a notably different flux. We use the X-ray temporal decay index $\alpha$ (defined such that $F_\text{x}(t)\propto t^{-\alpha}$; also retrieved from the UKSSDC online repository) to calculate the error due to the separation in time that does exist. This additional temporal error is determined using the formula \begin{equation}\label{eq:delta_B_ox}
    \Delta\B{ox} = |\alpha\log(1+dt_\%)|
\end{equation}
and then combined with the propagated uncertainty on $\B{ox}$. For each matched pair and resultant $\B{ox}$, we set a boolean flag for optical darkness according to the \citeauthor{2004ApJ...617L..21J} method if $\B{ox}<0.5$, and for the \citeauthor{2009ApJ...699.1087V} method if $\B{ox}<\B{x}-0.5$.

\subsection{Uncertainty handling \& error propagation}\label{sec:uncertainty}

All stages of the pipeline described above involve quantities with associated uncertainties. In many cases, this uncertainty is asymmetric (i.e., the error in the positive direction differs from the error in the negative direction). To ensure proper handling and propagation of all of these uncertainties, we utilize  the \texttt{asymmetric\_uncertainty} software package \citep{2022ascl.soft08005G}, a stand-alone Python library for representing such numbers. The package's main functionality lies in its implementation of a novel object type for representing quantities of the form $\mu_{-\sigma_-}^{+\sigma_+}$, where $\mu$ is the expected value and $\sigma_\pm$ are (not necessarily equivalent) uncertainties in the positive and negative direction. Instances of this class class behave appropriately under all standard mathematical operations (addition, division, exponentiation, etc.), and can be combined with one another (or other numerically-typed objects) using such operations to propagate their associated uncertainties. Mathematically, each object is treated as two conjoined and jointly normalized halves of a Gaussian probability distribution \citep[as introduced by][]{1982CSTM..11..879J}, with a PDF $P$ described by \begin{equation}\label{eq:splitnormal}
    P(x,\mu,\sigma_-,\sigma_+) = \frac{\sqrt{2}}{\sqrt{\pi}(\sigma_- + \sigma_+)}\cdot\begin{cases}\exp\left(-(x-\mu)^2/2\sigma_-^2\right) & x<\mu \\ \exp\left(-(x-\mu)^2/2\sigma_+^2\right) & x>\mu \end{cases}
\end{equation}
where $x$ is the independent random variable, $\mu$ is the $x$-coordinate of the peak, and $\sigma_\pm$ independently set the width to either side of that peak.

Notable examples of quantities involved in this analysis that have asymmetric uncertainties are the X-ray flux $F_\text{x}$, observation time $t_\text{x}$, and spectral index $\B{x}$. This means that each $\B{ox}$ ends up with an asymmetric uncertainty, since its calculation depends upon all of the aforementioned values. 
The software is also capable of handling classical (symmetric) uncertainties ($\sigma_+=\sigma_-$), as well as upper/lower limits (using variations on $\texttt{nominal}_{-\infty}^{+0}$ or $\texttt{nominal}_{-0}^{+\infty}$, respectively), making it a versatile computational tool.

\section{Results}\label{sec:results}

\begin{figure}
    \centering
    \includegraphics[width=0.9\linewidth]{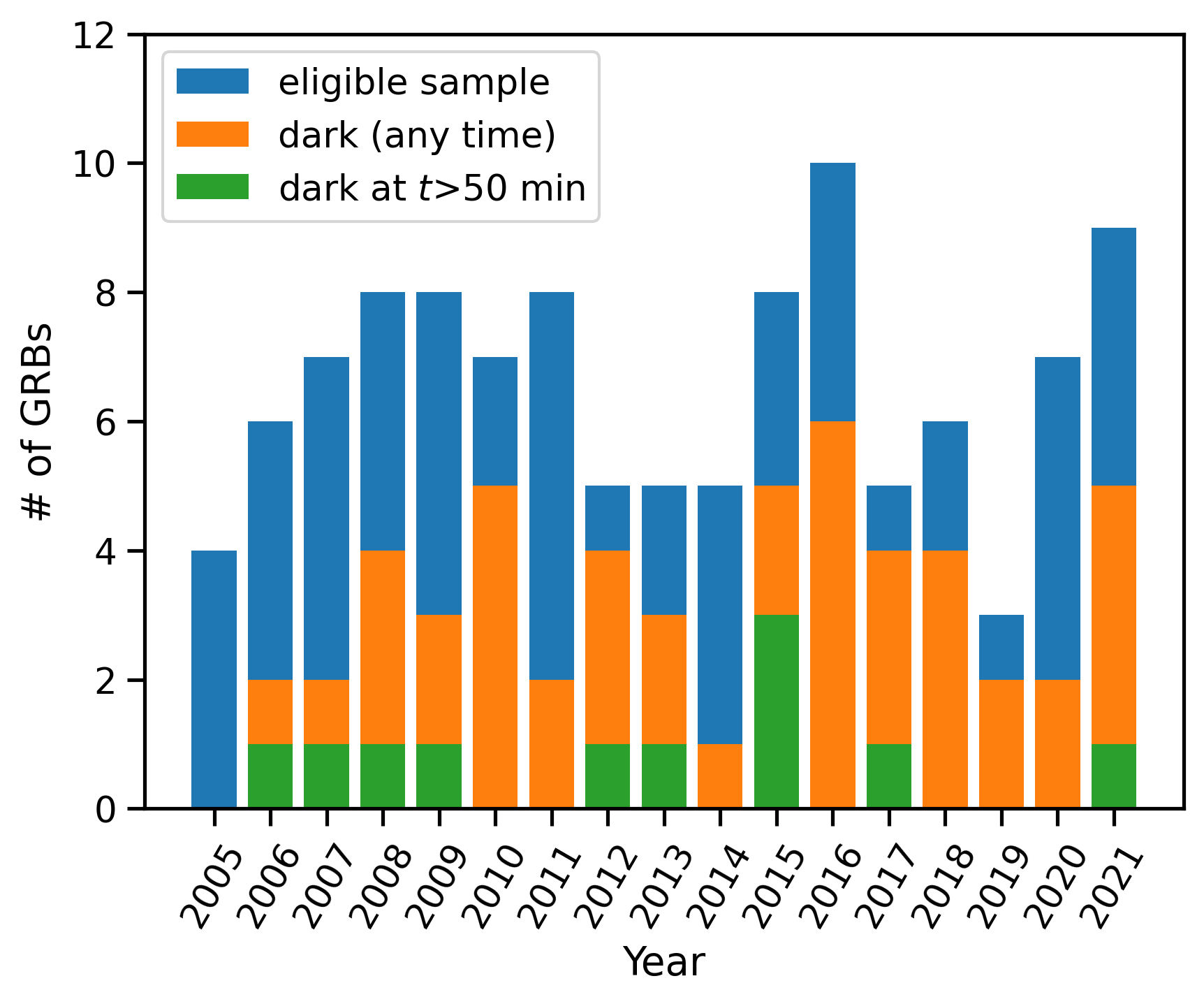}
    \caption{Bursts per year in our sample as a whole as compared to bursts per year that have at least one matched data pair that qualifies as dark by one or both methods.}
    \label{fig:per_year}
\end{figure}

The original sample of short \textit{Swift} GRBs consists of 193 events, spanning from February 2005 through the end of 2021. Of these, there is an X-ray light curve for 163 and at least one optical observation for 165. The overlap between these two sets (i.e., the number of bursts for which there are both X-ray and optical data) is 145. Of the latter, 108 bursts have at least one temporally matching set of data points with $dt_\%\leq0.2$, which comes out to $6.5\pm0.5$ candidate short GRBs per year of the sample. Finally, 54 bursts have at least one match that qualified as optically dark by at least one of the methods described above, yielding an average rate of $3.2\pm0.4$ empirically dark short bursts per year. These numbers give a rate of approximately $49\%\pm7\%$ of eligible bursts that are nominally dark (at some point) by one or both methods. The number of bursts per year is shown in Figure~\ref{fig:per_year}.

\begin{figure}
    \centering
    \includegraphics[width=0.9\linewidth, height=2in]{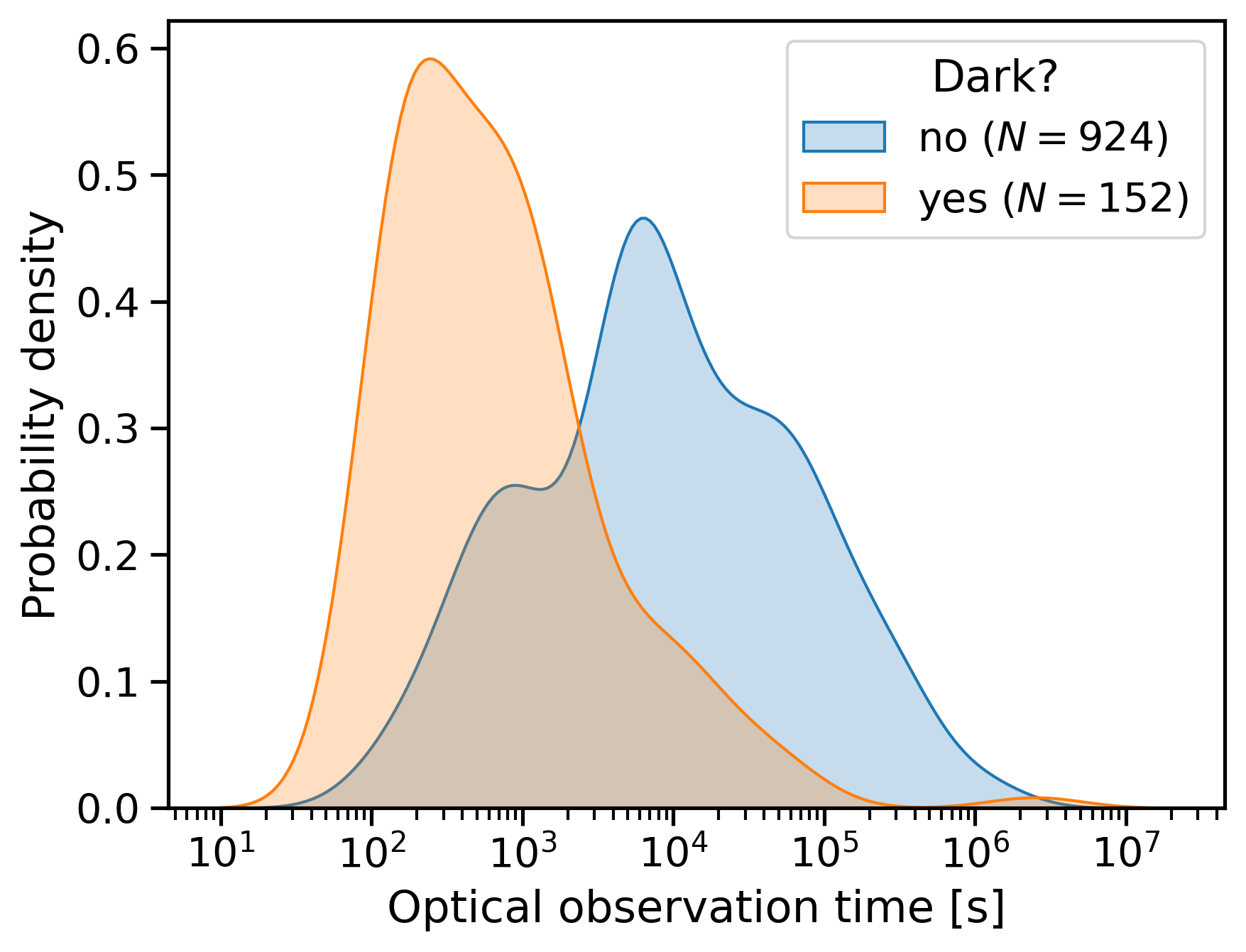}
    \caption{Kernel density plot showing the distribution of all optical observation times for dark data pairs and not dark data pairs. These distributions are not analogous, confirmed by a Kolmogorov-Smirnov (K-S) test statistic of $D=0.58$ with $p \approx 0$, which demonstrates with very high confidence that the two do not come from the same parent distribution.}
    \label{fig:obs_time_dist}
\end{figure}

However, the vast majority of these optically dark points come from very early-time follow-up observations. We can observe this effect in our sample of short GRBs as a whole by studying the time distributions of optical/X-ray observation pairs that qualify as dark versus not dark points. The number of bursts that qualify as dark, broken down into discrete classification of observation times ($<5$, $5-50$ and $>50$ minutes) and the breakdowns are given in Table~\ref{tab:contingency}, showing a clear trend between darkness classification and time after burst. Continuous distributions of observation times split by darkness classification are shown in Figure~\ref{fig:obs_time_dist}.

\begin{table}
    \centering
    \begin{tabular}{c|cc}
    \hline
    $t_\text{o}$  & Not dark  & Dark \\
    \hline
    <5 min    &  39 & 58 \\
    5--50 min & 205 & 68 \\
    >50 min   & 680 & 26 \\
    \hline
    \end{tabular}
    \caption{Darkness classification (by one or both methods) of temporally-matched observations across all GRBs, broken down by optical observation time, showing that early-time observations are disproportionately dark.}
    \label{tab:contingency}
\end{table}

To illustrate this effect quantitatively, we perform a search for events that have a calculated dark point at later times. We find 23 such bursts with a dark point beyond 5 minutes (300 seconds) after the start of the prompt gamma-ray emission, and only 10 with an optically dark point beyond 50 minutes (3000 seconds).

\begin{figure*}
    \centering
    \includegraphics[width=\linewidth]{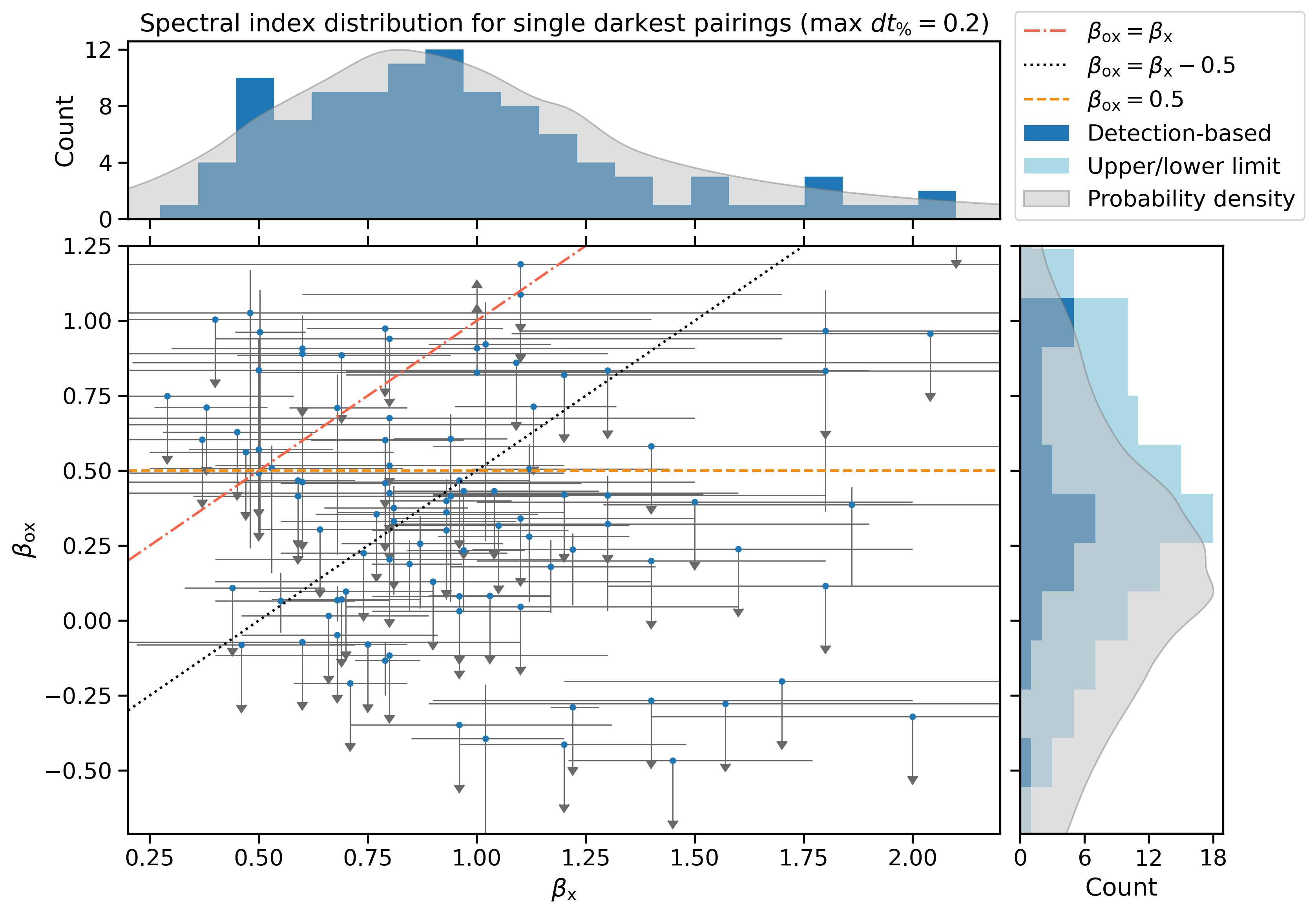}
    \caption{Each point in the plot above represents one unique burst and corresponds to the minimum value of $\B{ox}$ for that GRB. Lines are drawn at various $\B{x}$--$\B{ox}$ relations of interest (the \citeauthor{2009ApJ...699.1087V} definition of darkness is shown in dotted black, the \citeauthor{2004ApJ...617L..21J} criterion is the dashed orange line, and the dash-dotted red line indicates $\B{x}=\B{ox}$). Top and side plots show distributions for each variable created by summing asymmetric split-normal probability distribution functions based on the error bars of each point along the respective axis (see \S\ref{sec:uncertainty}).}
    \label{fig:beta_dist}
\end{figure*}

A complete summary of the resultant products of this work is visualized in Figure~\ref{fig:beta_dist}, which shows the distributions of $\B{x}$ and $\B{ox}$ relative to one another for the darkest (lowest value of $\B{ox}$) matched data pair for each GRB. This plot may therefore skew somewhat `dark', as it has not been corrected for the early-time anomalies discussed above. Interestingly, we note that the location of the apparent peak in the distribution of $\B{ox}$ lies just below $0.5$, which is \citeauthor{2004ApJ...617L..21J}'s cutoff for defining optical darkness.
 
\section{Discussion}\label{sec:discussion}

Our preliminary finding that just under half of short bursts exhibit optical darkness is unexpected, as previous studies \citep{2005ApJ...624..868R, 2011A&A...526A..30G, 2015MNRAS.449.2919L} have found a similar fraction for long GRBs. Due to the general differences in the environments and redshifts at which we expect short versus long GRBs to occur, it is surprising to find similar rates of optical darkness between the two populations. There are a number of possible reasons for this unexpected initial result.

\subsection{Early-time X-ray emission}\label{sec:early_xray}

\textit{Swift}'s short response time, in combination with the number of currently operational ground-based observatories capable of performing rapid follow up, often results in simultaneous observational coverage of GRB afterglows in the optical and X-ray bands. Because it is possible for afterglow emission in the X-ray and optical bands to fade over time at different rates \citep{2011A&A...526A..30G}, the resulting value of $\B{ox}$ can change over time as well, even over the course of a single burst's afterglow.\footnote{If the temporal decays of the optical and X-rays paralleled one another exactly, the two light curves would always be the same distance apart in logarithmic space, resulting in a constant value of $\B{ox}$.} This could be due to the optical and X-ray regimes being in different parts of the same broadband spectrum, or two different emission components contributing to the observed optical and/or X-ray emission.

\begin{figure*}
    \centering
    \includegraphics[width=0.49\linewidth]{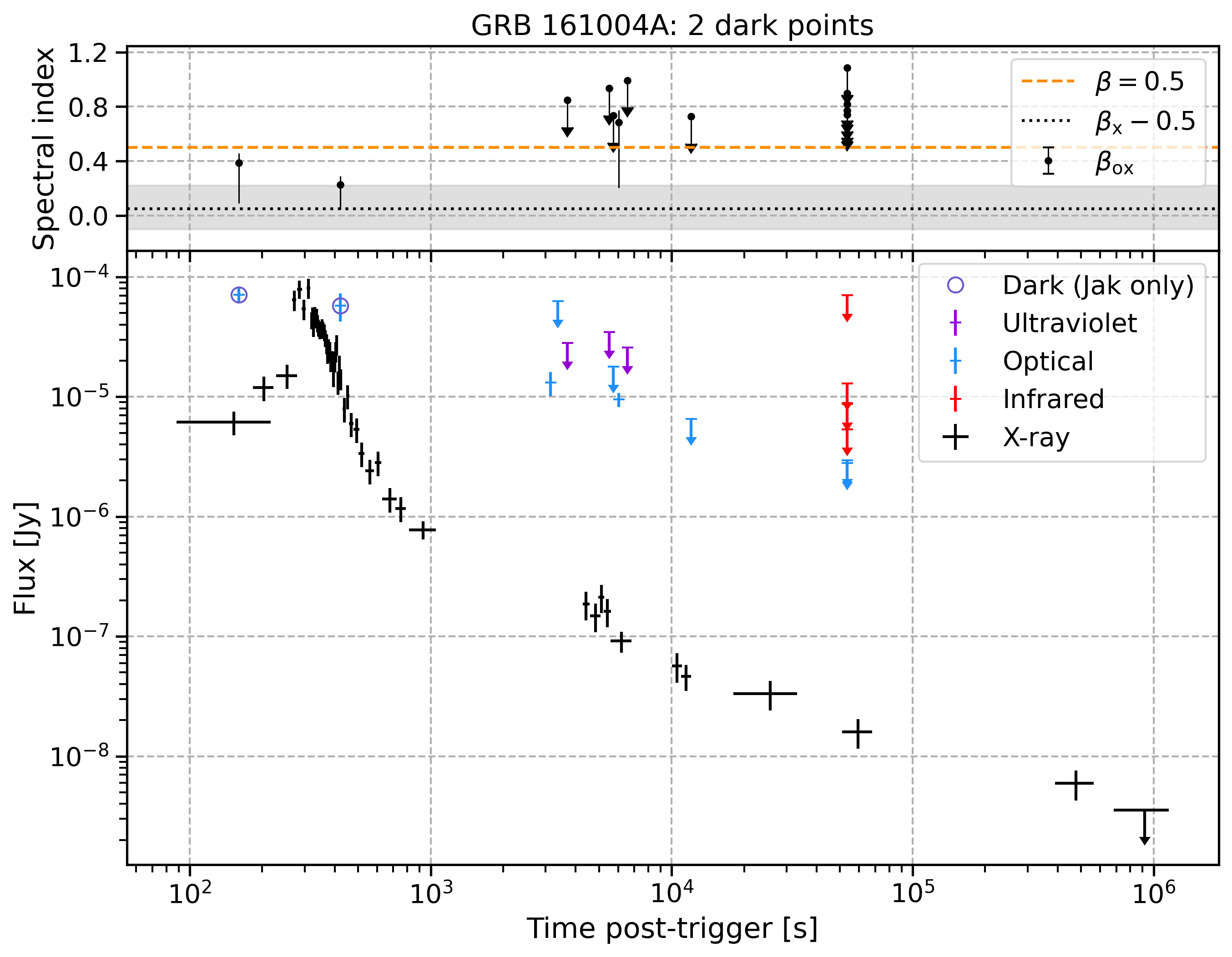}
    \includegraphics[width=0.49\linewidth]{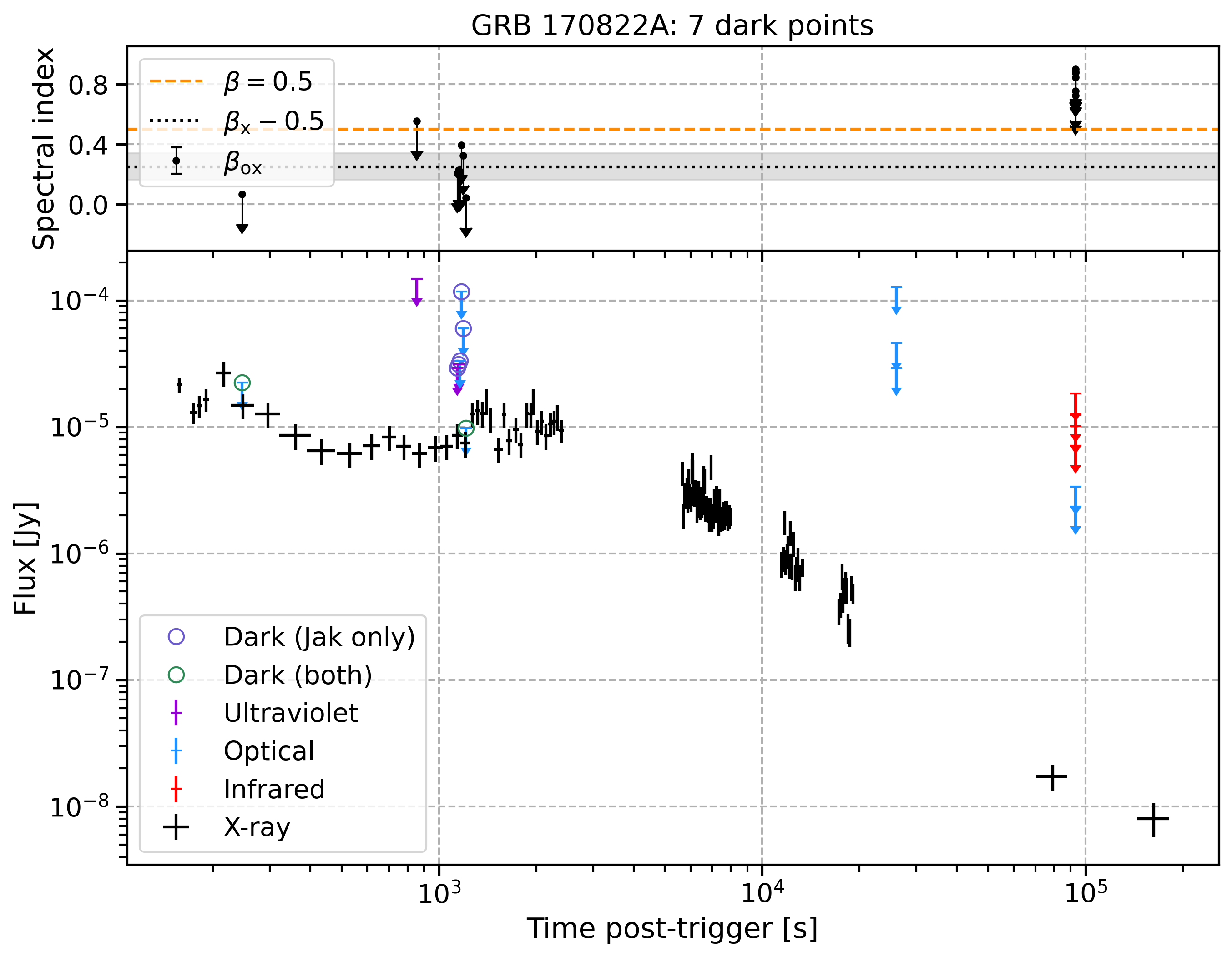}
    \caption{In the light curve of GRB\,161004A on the left, we note an obvious X-ray flare \citep[also flagged by the automated \textit{Swift}-XRT light curve fitting routine;][]{2007A&A...469..379E, 2009MNRAS.397.1177E} that peaks around $\delta t\approx300$ seconds, causing optical points at $\delta t=160$s and $\delta t\approx420$s to appear dark, even though the optical behavior looks fairly canonical. Similarly, we observe plateau-like behavior (and possibly a flare) evident in the light curve of GRB\,170822A on the right, lasting until around 2000 seconds post-trigger. Both are cases of anomalous X-ray behavior where it is clear that the light curve has not settled into `normal' decay, meaning that standard assumptions about the optical and X-ray regimes existing as two regions of the same broadband spectrum are not valid.}
    \label{fig:anomalous}
\end{figure*}

As discussed in \S\ref{sec:results}, we note numerous cases where a burst qualifies as dark at early times, but not at later times. To investigate the effects of observation time on perceived optical darkness, we inspected the light curves of individual dark bursts and noted that many such points where $\B{ox}<0.5$ or $\B{ox}<\B{x}-0.5$ (or both) coincide with times when there is clearly atypical behavior occurring in the X-rays. The canonical behavior for the temporal evolution of the X-ray afterglow \citep{2006ApJ...642..389N} is a brief period of very steep decay ($\alpha\approx3$), followed by a shallow decay ($\alpha\approx0.5$), and then finally a decay with an intermediate slope ($\alpha\approx1.0$--$1.5$). When the observed light curve behavior differs significantly from the latter, intermediate slope, our assumption about how the X-rays should behave is violated in comparison to the optical band, since there may be multiple emission components at play. Therefore, we hypothesize that many of these early points are dark by technicality, but not necessarily because of low optical flux; an X-ray excess is just as capable of causing $\B{ox}$ to be shallow. The lightcurves of GRBs 161004A and 170822A (Figure~\ref{fig:anomalous}) provide particularly notable examples of this, with an early X-ray flare and an extended plateau, respectively.

We conclude that although many bursts may qualify as optically dark in the numerical sense, it is often not in the interesting sense of the phenomenon that we seek. There are a number of possible reasons for this anomalous X-ray behavior at early times \citep{2006ApJ...642..389N,2006ApJ...646..351L}. There is a chance that early X-ray observations are catching the tail end of the burst's prompt emission or prolonged central engine activity. Regardless of whether the merger product is a rapidly spinning supra-massive neutron star or collapses immediately into a black hole, we expect that not all of the matter from the two progenitor objects will be consumed immediately: there is likely a short-lived accretion disk still actively fueling relativistic jets within the first few seconds or minutes after the burst. The current physical understanding of GRB afterglows is that emission arises from external shocks between the burst outflow and material in the surrounding environment. To explain the extra emission in the X~rays, we need an additional emission component beyond a standard forward shock afterglow model, and this could be the result of a number of theorized mechanisms, including prolonged central engine activity, stratified ejecta, or a reverse shock scenario. This is because there are also interactions within the jet structure that must be considered: for example, a faster-moving blast wave behind the main shock front may eventually catch up with it and inject additional energy. Once this has all played out, however, we observe the X-ray light curves settle into more typical behavior in time.

\subsection{Short GRBs with extended emission}\label{sec:extended}

\begin{figure}
    \centering
    \includegraphics[width=0.95\linewidth]{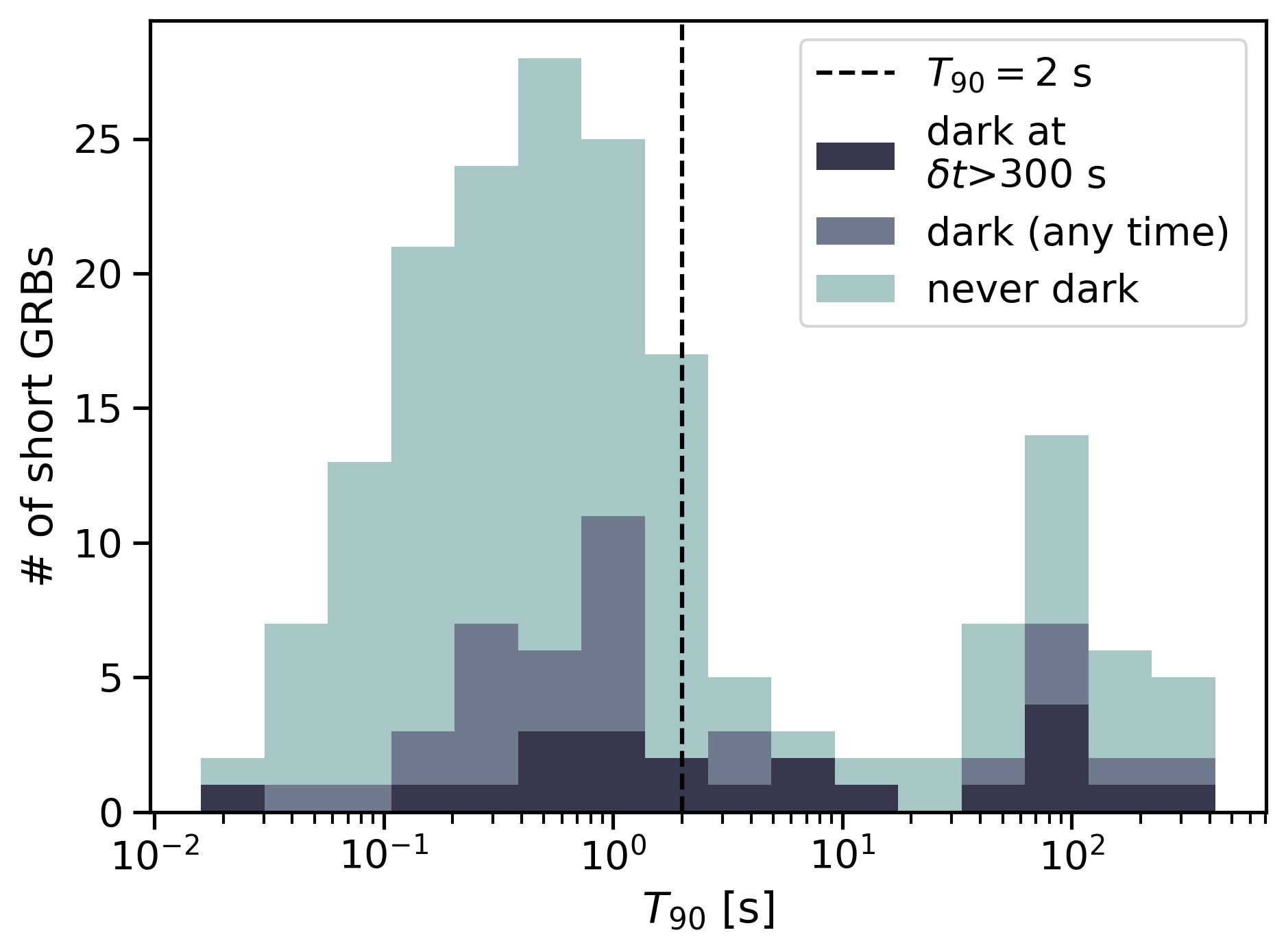}
    \caption{Histogram of $T_{90}$ values for bursts in our sample. The vertical line shows $T_{90}=2$s, the typically accepted value for defining the split between long- and short-duration GRBs. While the majority of events in our sample obey this, we also have some bursts that we deem to belong to the short class for other reasons and lie to the right of this dividing line. We note an apparent bimodality in this histogram, with the two peaks separated at $T_{90}\approx20$s.}
    \label{fig:EE_hist}
\end{figure}

Because of our sample selection methodology, there are a number of sample members that do not obey the $T_{90}\leq2$ second criteria. In fact, some have prompt durations on the order of tens or hundreds of seconds. These are bursts with extended emission \citep{2006ApJ...643..266N}, where high-energy emission continues beyond the main peak of the burst. An open question is what sets these bursts apart and why. Figure~\ref{fig:EE_hist} shows the distribution of our sample in $T_{90}$ space. Of note is the presence of two separate peaks that each appear to be distinct log-normal distributions, which reaffirms that our sample selection methodology is reasonable, and suggests the possibility that short bursts with extended emission (EE) form a distinct class with a different physical origin than typical short GRBs.

To probe whether these EE bursts arise from a different physical process, we examined the rates of optical darkness and late-time optical darkness on either side of the $T_{90}$ cutoff. We count the number of bursts on either side of this split and compare the relative fractions of bursts that possess optically dark data points as well as bursts with dark data points at late times (i.e., $\delta t>5$ minutes). If we find significantly different fractions of optical darkness or late-time optical darkness, we might infer that EE bursts (those in our sample but with comparatively large values of $T_{90}$) might be physically distinct from typical short bursts in their origins. The results of this analysis are shown in Table~\ref{tab:ee_split}. While these numbers may hint at EE bursts being more optically dark than non-EE short bursts, when we incorporate and propagate uncertainties (Poisson statistics; $\sigma_N\approx\pm\sqrt{N}$), the fractions' errors overlap at the 1.5--2$\sigma$ level, meaning no significant conclusion can be drawn.

A reanalysis of \textit{Swift}--BAT lightcurves by \citet{2021ApJ...911L..28D} found that a majority of sGRBs at high redshift ($z\gtrsim1$) display extended emission, despite often having a measured $T_{90}$ of $\leq2$~s. They note the possibility that this is due to EE bursts arising from progenitors other than binary neutron star mergers. Other studies \citep{2008MNRAS.385L..10T, 2020ApJ...895...58G} have suggested that EE short bursts are the result of neutron star-black hole binary mergers, as opposed to mergers of binary neutron stars. This is supported by physical modeling, as well as the fact that EE bursts are typically found closer to their host galaxies than non-EE bursts \citep{2022MNRAS.515.4890O}. This is notable because if true, it could reasonably be expected to contribute to optical darkness as well. Although our results hint at this possibility, larger number statistics are required to support or refute this.

\begin{table}
    \centering
    \begin{tabular}{cccc}
        $T_{90}$ & dark at $\delta t>5$~min & dark at any time  & entire sample \\ \hline
        $>20$ s & 7 ($22 \pm 9$ \%) & 13 ($41 \pm 13$ \%) & 32 (100\%) \\
        $\leq20$ s & 15 ($10 \pm 3$ \%) & 38 ($26 \pm 5$ \%) & 149 (100\%) \\
        \hline
        $>2$ s & 11 ($22 \pm 7$ \%) & 19 ($38 \pm 10$ \%) & 50 (100\%) \\
        $\leq2$ s & 11 ($8 \pm 3$ \%) & 32 ($24 \pm 5$ \%) & 131 (100\%) \\
    \end{tabular}
    \caption{Number of optically dark GRBs, with (in parentheses) the fraction of the row total that each entry represents. Of note would be a rate that differs significantly for bursts above one of the $T_{90}$ values versus below it. We show the breakdown using two different cutoff durations for defining EE bursts: $T_{90}=2$ seconds, which is the classical criterion used for defining short bursts, and $T_{90}=20$ seconds, which is the apparently more natural split in Fig.~\ref{fig:EE_hist}.}
    \label{tab:ee_split}
\end{table}

\subsection{Meaningfully dark bursts}

The early-time effect (discussed in \S\ref{sec:early_xray}) is so widespread that when we correct for it, we find only 4 bursts in our sample that exhibit meaningful optical darkness: GRBs 060121, 090423, 130603B, and 170728B.

\subsubsection{GRB 060121}

\begin{figure}
    \centering
    \includegraphics[width=0.95\linewidth]{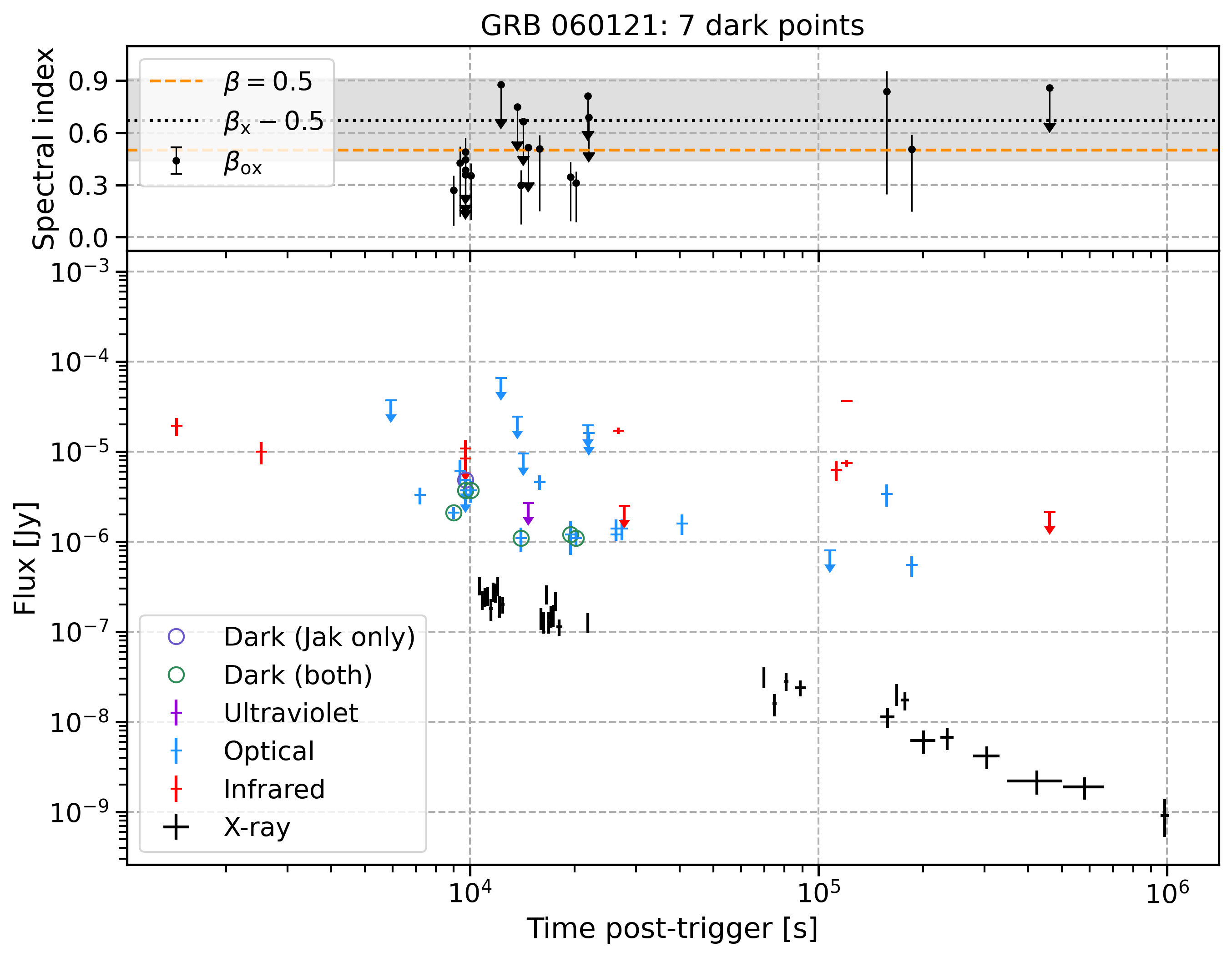}
    \caption{Multi-wavelength afterglow light curve of GRB\,060121, showing optical darkness between approximately $1\cdot10^4$ and $2\cdot10^4$ seconds.}
    \label{fig:060121}
\end{figure}

We find that GRB\,060121 attains a minimum optical-to-X-ray spectral index of $\B{ox}\cong0.18^{+0.09}_{-0.15}$ at $\delta t\approx3.9$ hr (see Figure~\ref{fig:060121}). Its most likely redshift is $z\sim4.6$, with a possibility that it might be $z\sim1.7$ \citep{2006ApJ...648L..83D}. Either way, this qualifies it as fairly high-redshift, especially for a short GRB. The Lyman\nobreakdash-$\alpha$ forest, caused by clouds of neutral hydrogen in the intergalactic medium, can cause an extra absorption in the optical regime, possibly explaining the optical darkness of this GRB.
However, if it is indeed at the higher redshift of $\sim$4.6, corresponding to a Universe that is $<2$ Gyr old, a BNS merger is an unlikely progenitor of this burst, and it may have resulted from a collapsar instead. This assumes a typical delay time for BNS mergers of 3--4 Gyr, as found by \citet{2015MNRAS.448.3026W} and \citet{2020A&A...634L...2S}. However, \citet{2019MNRAS.486.2896S} have suggested a much faster timescale for BNS mergers, with an average coalescence time of 300--500 Myr. In this case, a merger-induced GRB is feasible at $z=4.6$.
Further compounding the uncertainty surrounding the origins of GRB\,060121, \citet{2021ApJ...911L..28D} find evidence that despite its prompt duration of $T_{90}=1.97\pm0.06$ seconds, its lightcurve also exhibited extended emission, possibly hinting at a non-BNS progenitor as well.

\subsubsection{GRB 090423}

\begin{figure}
    \centering
    \includegraphics[width=0.95\linewidth]{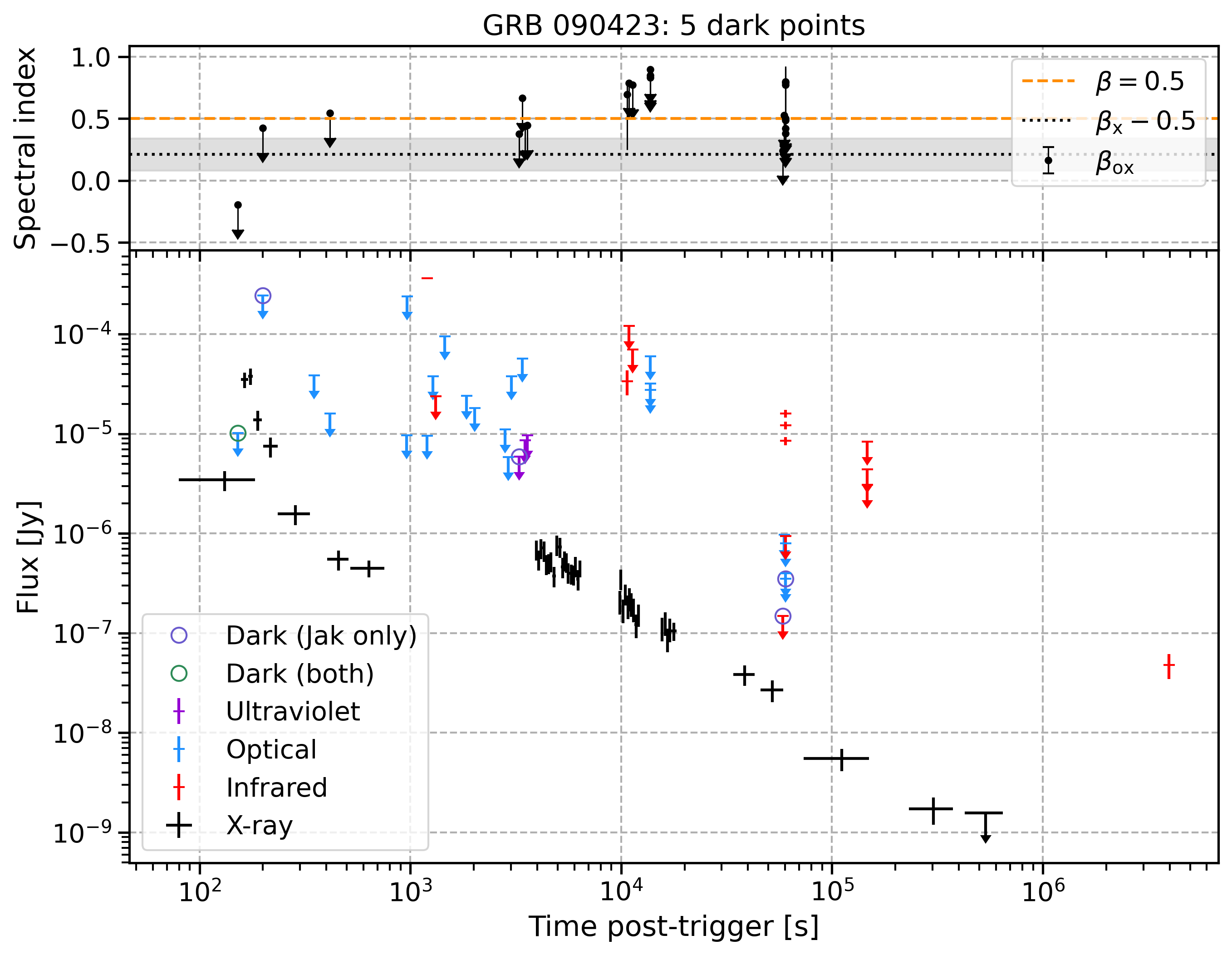}
    \caption{This light curve shows very early-time optical darkness in GRB\,090423, possibly resulting from an apparent X-ray flare that peaks between $100$ and $200$ seconds. It is also dark at later times: in UV at $3\cdot10^3$s, and in optical/nIR at $6\cdot10^4$ seconds.}
    \label{fig:090423}
\end{figure}

GRB\,090423 was and is one of the highest-redshift GRBs ever detected, at $z\sim8.2$ \citep{2009Natur.461.1254T, 2009Natur.461.1258S}. Using a standard cosmological model \citep{2014ApJ...794..135B}, this $z$ corresponds to a cosmological age of only about 600 million years---over 13 billion years ago. Though it is darkest at early times ($\B{ox}<-0.21$ at $\delta t=152$ s), the dark point at $\delta t\approx55$ min is an upper limit of $\B{ox}<0.38$ (see Figure~\ref{fig:090423}).
When the GRB was first detected, its initial classification as long or short was inconclusive. While the $T_{90}$ duration ($10.3\pm1.1$ seconds in the observer frame, $1.1\pm0.1$ seconds in the GRB rest frame), spectral lag, and peak energy were consistent with a short burst \citep[][explaining why it is present in our sample]{2009GCN..9241....1K, 2009ApJ...703.1696Z}, it has since been confirmed to be a high-redshift long GRB, which explains its optical darkness: at a redshift of $z\sim8.2$, the Lyman-$\alpha$ forest would span wavelengths from 121.6 nm to 1119 nm, i.e., the entire optical regime.

\subsubsection{GRB 130603B}

\begin{figure}
    \centering
    \includegraphics[width=0.95\linewidth]{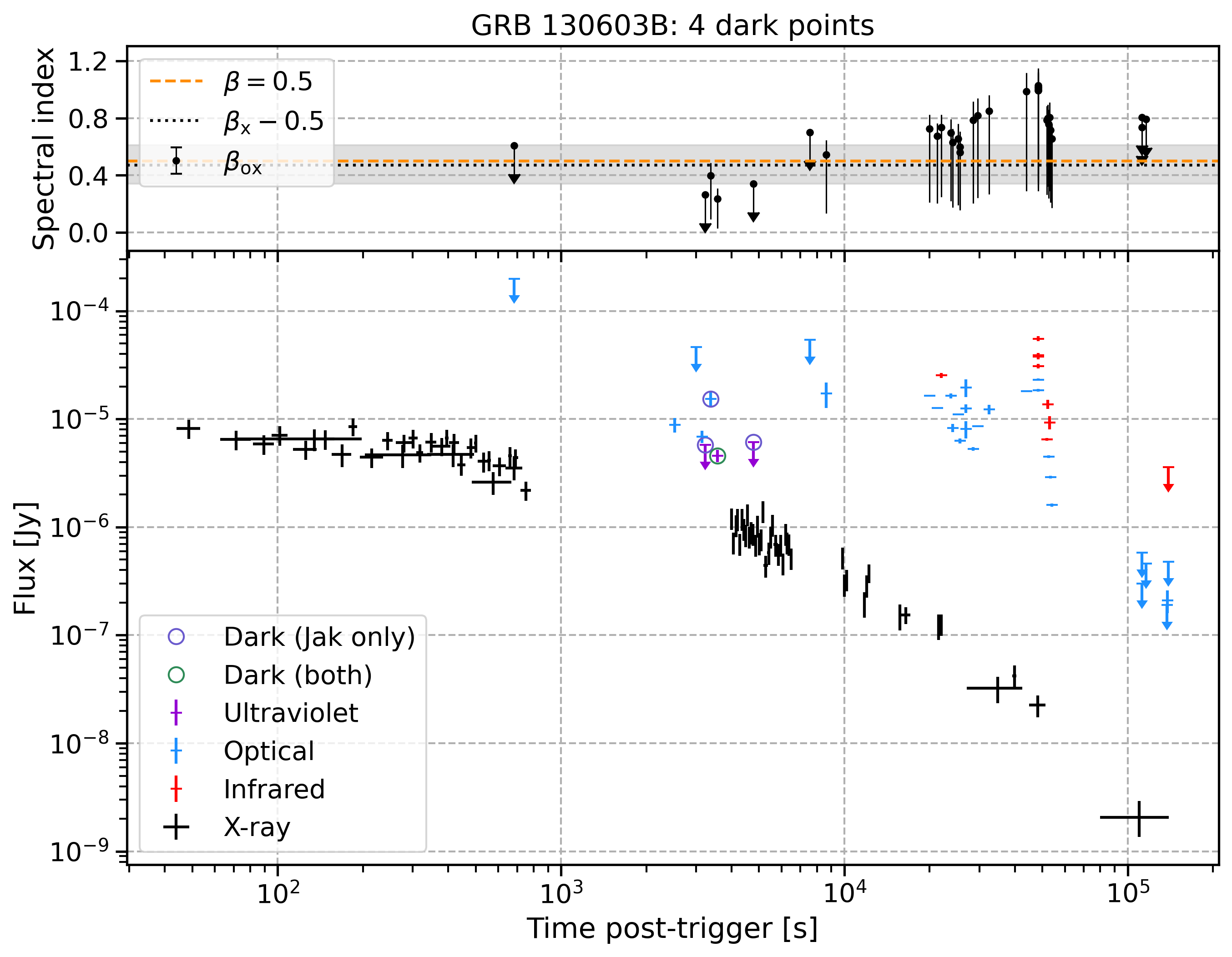}
    \caption{For GRB\,130603B, we identify a cluster of optically dark data points around $\sim$1~hour, around the time that \citeauthor{2014ApJ...780..118F}'s magnetar model implies that excess X-ray emission becomes a significant contributor.}
    \label{fig:130603B}
\end{figure}

With $T_{90}=0.18\pm0.02$ seconds, GRB\,130603B lies solidly in the short class of $\gamma$-ray bursts. Spectroscopic analysis by \citet{2014A&A...563A..62D} assigns a redshift of $z\sim0.36$, ruling out a high redshift as a viable explanation for optical darkness in this case. Of interest, however, is the late-time excess X-ray emission noted by \citet{2014ApJ...780..118F}, which they attribute to a rapidly spinning supra-massive magnetar merger product. \citet{2014MNRAS.438..240G} supports the millisecond magnetar as a candidate central engine capable of producing both extended emission and a later-time X-ray plateau, with the extra emission powered by rotational spin-down. \citeauthor{2014ApJ...780..118F} found that both the late-time ($\gtrsim 3\cdot10^3$ seconds) spectrum and the light curve of GRB\,130603B were consistent with this model.
In its light curve (Figure~\ref{fig:130603B}), we note that our optically dark points occur around this time as well: we find a minimum $\B{ox}=0.23_{-0.21}^{+0.07}$ at approximately 1~hour post-trigger. This indicates that the X-ray excess due to central engine activity may indeed be to blame for the shallow values of $\B{ox}$.

\subsubsection{GRB 170728B}

There is little published work on GRB\,170728B. It is optically darkest at $t-t_0=16$ minutes with $\B{ox}=-0.13_{-0.11}^{+0.06}$, but we also find an upper limit of $\B{ox}<-0.014$ at $\sim3.5$ hours. With a short, multi-peaked burst structure but a $T_{90}$ of $47.7\pm25.2$ seconds \citep{2017GCN.21384....1U}, this burst could be considered a short burst with extended emission, like those previously discussed in \S\ref{sec:extended} \citep[and is identified as such by][]{2022ApJ...940...57N}. \citet{2022ApJ...940...56F} propose a host galaxy association with a spectroscopic redshift of $z=1.272$. Optical observations of GRB\,170728B's afterglow that satisfy the criteria for optical darkness are available at a range of wavelengths across the optical range, with a majority of them \citep{2017GCN.21394....1D} being in the $R_c$ band ($\lambda_\text{eff}\approx636$\,nm). This rules out redshift as a viable explanation for optical darkness in this case.

\begin{figure}
    \centering
    \includegraphics[width=0.95\linewidth]{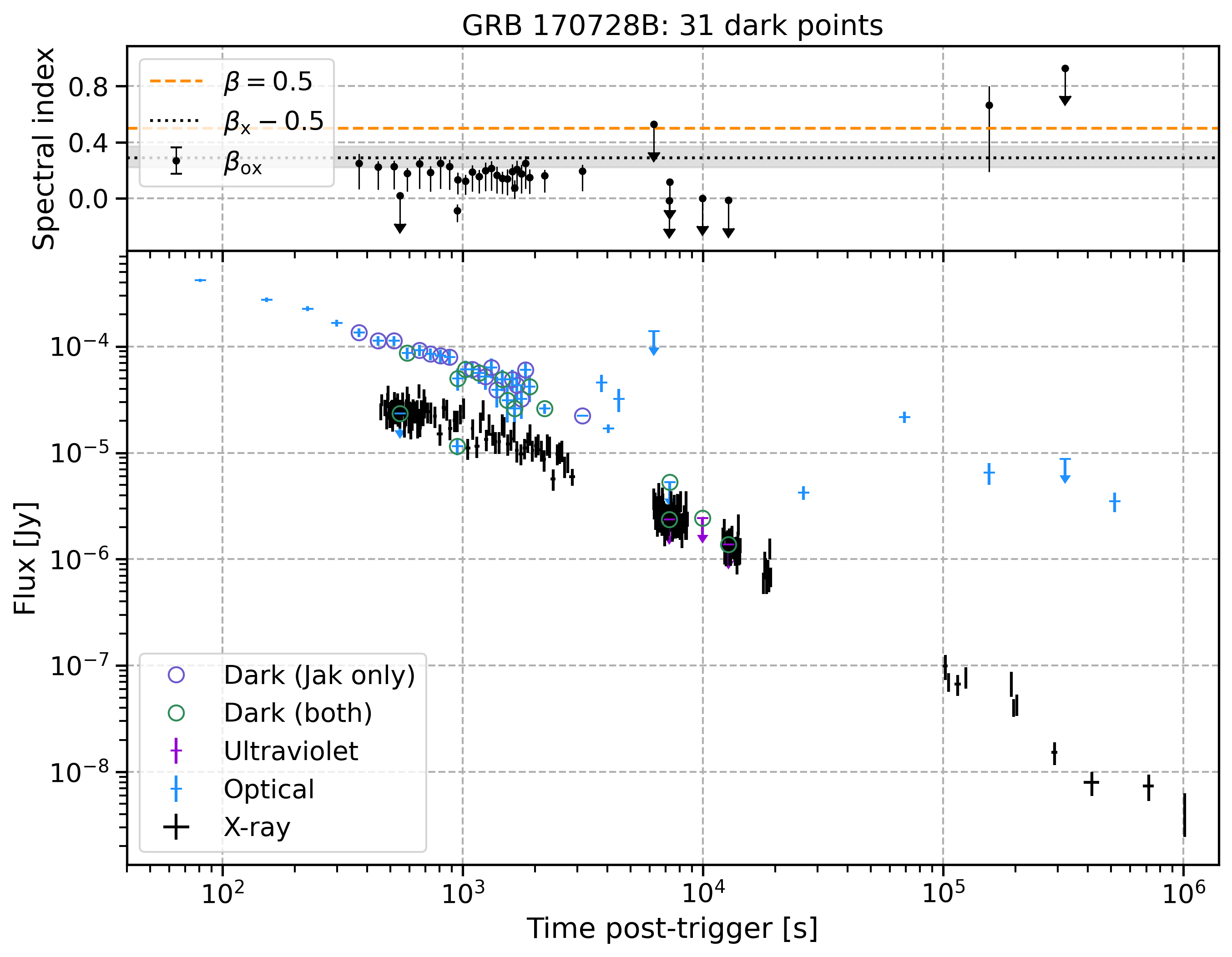}
    \caption{While the majority of the points that qualify GRB~170728B as optically dark occur at early times, there is also a cluster around $10^4$ seconds.}
    \label{fig:170728B}
\end{figure}

\section{Conclusions}\label{sec:conclusion}

We present a complete, systematic study of optical darkness in short GRBs, a phenomenon that has until now only been studied extensively in long GRBs. To this end, we present our complete, scalable, and largely automated software pipeline, as well as a comprehensive catalog of short GRBs that comprises >3,000 optical observations and >5,500 X-ray data points from nearly 200 individual bursts.

Previous work on long GRBs has found a rate of optical darkness around half \citep[][and references therein]{2015MNRAS.449.2919L}. While our initial result for short GRBs is consistent with this, many instances of optical darkness in our sample stem from early-time observations, and we determine that in most cases an excess of X-ray emission is to blame rather than any significant optical deficit. The classification of data points as optically dark or not depends heavily on how quickly follow-up observations are obtained, and assessing optical darkness using existing criteria is better done at later times once X-ray emission has settled into regular decay. The high rate of occurrence of anomalous early-time X-ray behavior suggests that the standard assumption of a purely synchrotron afterglow with one broadband emission component is not a complete picture. While previous studies of optical darkness in long GRBs have avoided this problem by using data obtained at later times, the comparatively faint overall nature of short GRB afterglows means that the availability of data skews earlier and identifying true optical darkness is more difficult.

When we account for early-time effects, we find that, as expected, optical darkness is much more rare in short GRBs than in long ones. We identify only 4 of our GRBs that are optically dark after the X-ray lightcurve has entered regular decay, and one of them is actually a long GRB. Because our eligible sample (of bursts with temporally-matched data) consists of 108 GRBs, this number represents a \textit{true} optical darkness rate of less than 3\%. To explain the optical darkness in these few individual cases, we turn to redshift, late-time X-ray excess, or the possibility of heterogeneity in short GRB progenitors, remnants, and environments that could cause discrepant afterglow behavior.

The tools and results presented here are structured so as to make updating and keeping the catalog up to date as straightforward as possible. In addition to newly developed computational tools with wide-ranging cross-disciplinary applicability, this work provides a robust framework for further investigation and analysis of optical darkness in both long and short GRBs.

\section*{Acknowledgements}

The authors thank the anonymous referee for their thoughtful and constructive feedback that helped strengthen this work. The authors would also like to thank Brendan O'Connor for useful discussions and feedback on this manuscript. This work made use of data supplied by the \href{https://www.swift.ac.uk/xrt_products/index.php}{UK Swift Science Data Centre} at the University of Leicester. This research has made use of the \href{http://svo2.cab.inta-csic.es/theory/fps/}{SVO Filter Profile Service} supported from the Spanish MINECO through grant AYA2017-84089. This research made use of \href{http://www.astropy.org}{Astropy}, a community-developed core Python package for Astronomy \citep{astropy:2013, astropy:2018}.

\section*{Data Availability}

The complete datasets underlying this article are available on GitHub, in the repository \href{https://github.com/cgobat/dark-GRBs}{cgobat/dark-GRBs}. The data were derived from sources in the public domain: \href{https://www.swift.ac.uk/}{UKSSDC}, the \href{https://gcn.gsfc.nasa.gov/}{GCN Circulars Archive}, and previously published works \citep[][among others]{2015ApJ...815..102F, 2021ApJ...916...89R}.



\bibliographystyle{mnras}
\bibliography{main}




\appendix

\section{Supplementary Tables}

Table~\ref{tab:summary} contains a summary of the available data, matches, and optical darkness results for each GRB in our sample.

\begin{table}
    \centering
    \begin{tabular}{cccccc}
        \hline
            GRB & $T_{90}$ [sec] & X-ray & Optical & Temporal matches & Dark \\
        \hline
        211227A &  83.792 &   101 &      22 &               86 &   83 \\
        211106A &    1.75 &     4 &      13 &                1 &    0 \\
        211023B &   1.296 &    19 &      39 &               32 &    2 \\
        210919A &   0.164 &     2 &      35 &                2 &    0 \\
        210726A &   0.388 &    15 &       5 &                2 &    1 \\
            ... &     ... &   ... &     ... &              ... &  ... \\
         050906 &   0.064 &     1 &       1 &                0 &    0 \\
         050813 &   0.448 &     2 &      12 &                0 &    0 \\
         050724 &  95.964 &   238 &       7 &               55 &    0 \\
        050509B &   0.048 &     3 &       1 &                0 &    0 \\
         050202 &   0.112 &     0 &       1 &                0 &    0 \\
        \hline
        \end{tabular}
    \caption{Number of available data points for each of the short GRBs in our sample, along with the number of temporally-matching points and the resulting number of points that qualify as `dark' by one or both of the two methods described above. \\    
    Only a portion of this table is shown here to demonstrate its form and content. A machine-readable version of the full table is available.}
    \label{tab:summary}
\end{table}


\bsp 
\label{lastpage}
\end{document}